# F1000 recommendations as a new data source for research evaluation: A comparison with citations

Ludo Waltman and Rodrigo Costas

Centre for Science and Technology Studies, Leiden University, The Netherlands

{waltmanlr, rcostas}@cwts.leidenuniv.nl

F1000 is a post-publication peer review service for biological and medical research. F1000 aims to recommend important publications in the biomedical literature, and from this perspective F1000 could be an interesting tool for research evaluation. By linking the complete database of F1000 recommendations to the Web of Science bibliographic database, we are able to make a comprehensive comparison between F1000 recommendations and citations. We find that about 2% of the publications in the biomedical literature receive at least one F1000 recommendation. Recommended publications on average receive 1.30 recommendations, and over 90% of the recommendations are given within half a year after a publication has appeared. There turns out to be a clear correlation between F1000 recommendations and citations. However, the correlation is relatively weak, at least weaker than the correlation between journal impact and citations. More research is needed to identify the main reasons for differences between recommendations and citations in assessing the impact of publications.

## 1. Introduction

Assessing the quality or impact of scientific outputs is one of the major challenges in research evaluations. The two most commonly used instruments for assessing scientific quality or scientific impact are peer review (Bornmann, 2011) and citation analysis (Moed, 2005; Nicolaisen, 2007). Both instruments have their own strengths and weaknesses. Citation analysis can be applied at a large scale without too much effort, but the number of citations received by a publication is determined by a large variety of factors (Bornmann & Daniel, 2008), only some of which reflect a publication's quality or impact. Citation analysis may also be vulnerable to gaming, for instance by researchers who change their publication and citation practices in order to be assessed more favorably by citation-based impact measures. Compared with citation analysis, peer review is usually seen as a more trustworthy approach to assessing scientific quality, but at the same time the literature suggests that peer review judgments may be influenced by various types of biases (Bornmann, 2011). A practical problem of peer review also is that it can be quite expensive and time consuming to undertake. Given the strengths and weaknesses of peer review and citation analysis, it is often recommended to use both instruments in a combined fashion (e.g., Butler, 2007; Moed, 2007). This is indeed the approach that is taken in many research evaluations.

The recent introduction of so-called 'altmetrics' (Priem & Hemminger, 2010; Priem, Piwowar, & Hemminger, 2012; Priem, Taraborelli, Groth, & Neylon, 2010) may lead to the development new instruments for research evaluation. Altmetrics refers to data sources, tools, and metrics (other than citations) that provide potentially relevant information on the impact of scientific outputs (e.g., the number of times a publication has been tweeted, shared in Facebook, or read in Mendeley). Altmetrics opens the door to a broader interpretation of the concept of impact and to more diverse forms of impact analysis. At the same time, it has been argued that altmetrics still needs to overcome important problems in order to become a robust and stable instrument for research evaluation (Wouters & Costas, 2012).

Among the various altmetrics tools, there is one that deserves special attention, particularly because of its innovative use of peer review. This is Faculty of 1000, abbreviated as F1000 and recently renamed as F1000Prime (see http://f1000.com/prime). F1000 is a commercial online post-publication peer review service for biological and medical research. It was launched in 2002,[1] and so far it has collected reviews of over 100,000 biomedical publications. Reviews are produced by more than 5000 peer-nominated researchers and clinicians, referred to as F1000 faculty members. Faculty members are requested to select the most interesting publications they read and to provide reviews of these publications. A review of a publication consists of a recommendation ('good', 'very good', or 'exceptional') along with an explanation of the strengths and possibly also the weaknesses of the publication. Faculty members can choose to review any primary research article from any journal, without being limited to recent publications or publications indexed in PubMed. From a research evaluation point of view, F1000 is a quite unique service, offering peer review judgments on individual publications in a large-scale, systematic, and mixed qualitative and quantitative fashion, with reviews being available to anyone with a subscription to the service.

---

[1] In 2002, F1000 was referred to as F1000 Biology. F1000 Medicine was launched in 2006. Later on, the two services were combined.

In this paper, we present a large-scale analysis of F1000 recommendations, focusing in particular on comparing recommendations with citations. Our analysis aims to provide insight into the potential value of F1000 recommendations for research evaluation purposes. We are interested to see, for instance, to what extent recommendations correlate with citations, whether recommendations can be regarded as predictors of citations, or whether recommendations perhaps capture a different type of impact than citations do. F1000 recommendations have been studied before (Allen, Jones, Dolby, Lynn, & Walport, 2009; Bornmann & Leydesdorff, 2013; Li & Thelwall, 2012; Medical Research Council, 2009; Mohammadi & Thelwall, in press; Priem et al., 2012; Wardle, 2010; Wets, Weedon, & Velterop, 2003), but earlier studies were all based on relatively small data sets. In the present study, F1000 recommendations are analyzed in a much more comprehensive manner.

This paper also contributes to the literature on the relationship between citations and peer review (Bornmann, 2011; Moed, 2005; Nederhof, 1988). This relationship has been extensively studied, but there is a lack of large-scale comparisons between citations and peer review at the level of individual publications. Our analysis offers such a comparison.

The rest of this paper is organized as follows. In Section 2, we discuss the data that we use in our analysis as well as our methodology for processing the data. In Section 3, we present the results of our analysis. And finally, in Section 4, we summarize our main conclusions.

## 2. Data and methodology

In January 2013, F1000 provided us with data on all 132,844 recommendations made in their system at that time. For each recommendation, we received a score (1 = 'good'; 2 = 'very good'; 3 = 'exceptional'), the date at which the recommendation was given, and some bibliographic data on the publication being recommended. Of the 132,844 records, 182 actually cannot be regarded as true recommendations. These 182 records, which do not have a recommendation score, represent dissenting opinions, that is, cases in which an F1000 faculty member indicates that he or she does not agree with a recommendation given by another faculty member (see also http://f1000.com/prime/about/whatis). We excluded the 182 records from our analysis. Hence, our analysis includes 132,662 recommendations. These recommendations relate to 102,360 unique publications, which means that the average number of recommendations per publication equals 1.30 (taking into account only publications with at least one recommendation).

It should be mentioned that some recommendations have been given in a special way. Normally, F1000 faculty members read publications and if they consider a publication of sufficient interest, they may choose to recommend it. However, there is another way in which recommendations can be given. F1000 publishes two open access review journals: *F1000 Reports Biology* and *F1000 Reports Medicine* (see http://f1000.com/prime/reports). Authors of the review articles in these journals may add a recommendation to some of the publications they cite. These recommendations are included in the F1000 system in the same way as ordinary recommendations. They are also included in the data set that we have received from F1000. In this data set, it is not possible to distinguish the special recommendations from the ordinary ones, and our analysis therefore simply includes both types of recommendations. It has been suggested to us by F1000 that, in comparison with ordinary recommendations, recommendations originating from review articles in *F1000 Reports Biology* and *F1000 Reports Medicine* may tend to be given to older publications, but we have not been able to verify this ourselves.

Based on the bibliographic data provided by F1000, we linked the publications in the F1000 data set to publications in Thomson Reuters' Web of Science (WoS) database. A link between a publication in the F1000 data set and a publication in the WoS database was established if the publications had either the same digital object identifier (DOI) or the same journal title, volume number, issue number, and first author name (i.e., last name and initials). Perfect matches on journal title, volume number, issue number, and first author name were required, although we did perform some manual cleaning of the most common journal titles in the F1000 data set. Overall, there are 95,385 publications for which a link could be established between the F1000 data set and the WoS database. This corresponds with a matching rate of 95,385 / 102,360 = 93.2%. The 95,385 matched publications have been recommended 124,320 times. We note that our procedure for matching publications is quite conservative. We therefore expect there to be almost no incorrect matches. A less conservative matching procedure would have produced more matches, but most likely it would also have resulted in significantly more errors.

The first part of our analysis (presented in Subsections 3.1 and 3.2), which reports some more general statistics on F1000 recommendations, is based on the entire F1000 data set. The second part of our analysis (presented in Subsections 3.3, 3.4, and 3.5), which focuses mainly on the comparison between recommendations and citations, is based on a more restricted data set. For the purpose of the comparison between recommendations and citations, we restrict our analysis to publications from the period 2006–2009 and we include only publications of the WoS document types *article* and *review*. Also, for consistency with the way in which we count citations (see below), we only take into account recommendations given in the year in which a publication appeared or in one of the next two years. There turned out to be 38,369 publications that satisfy our criteria and that have at least one recommendation. For each of these publications, we determined the 'microfield' to which the publication belongs in the publication-level classification system of science recently developed by one of us (Waltman & Van Eck, 2012). This classification system includes 22,412 microfields, each consisting of at least 50 and at most a few thousand publications from the period 2001–2010. Of the above-mentioned 38,369 publications, 42 turned out not to be included in the classification system. The remaining 38,327 publications were found to belong to 5,908 different microfields. The overall number of publications in these 5,908 microfields in the period 2006–2009 is 1,707,631. Our comparison between recommendations and citations is based on these 1.7 million publications.

For each of the 1.7 million publications, we counted the number of citations received within a three-year citation window (i.e., in the year in which the publication appeared and in the next two years). Hence, citations were counted within the same time window as recommendations, so that we can make a fair comparison between the two. We also determined a journal citation score for each publication. The journal citation score of a publication in journal X equals the average number of citations received by all publications in journal X in the period 2006–2009. In the calculation of journal citation scores, only publications of the WoS document types *article* and *review* were considered. Citations were again counted within a three-year citation window.

## 3. Results

The presentation of the results of our analysis is split into five subsections. We first provide some general statistics on F1000 recommendations (Subsection 3.1). We then discuss the issues of the timing of recommendations (Subsection 3.2) and of the recommendation activity per field of science (Subsection 3.3). Finally, we extensively compare F1000 recommendation with citations (Subsections 3.4 and 3.5).

**3.1. General statistics**

We first present some general statistics on F1000 recommendations. Of the 132,662 recommendations, 77,674 (58.6%) have a score of 1 ('good'), 45,889 (34.6%) have a score of 2 ('very good'), and 9,099 (6.9%) have a score of 3 ('exceptional'). Hence, F1000 faculty members seem quite careful with the 'exceptional' judgment, as they use it in less than 7% of their recommendations.

As shown in Figure 1, the first recommendations were given in 2001. There has been an increasing trend in the number of recommendations given per year, with more than 16,000 recommendations given in 2012. A significant increase in the yearly number of recommendations took place between 2005 and 2006, which coincides with the launch of F1000 Medicine. Figure 2 indicates that for each of the three scores the proportion of recommendations has been more or less stable over time.

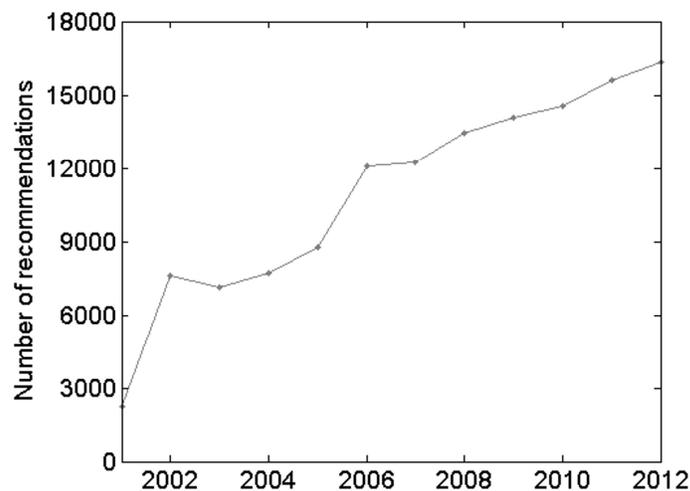

Figure 1. Number of recommendations per year.

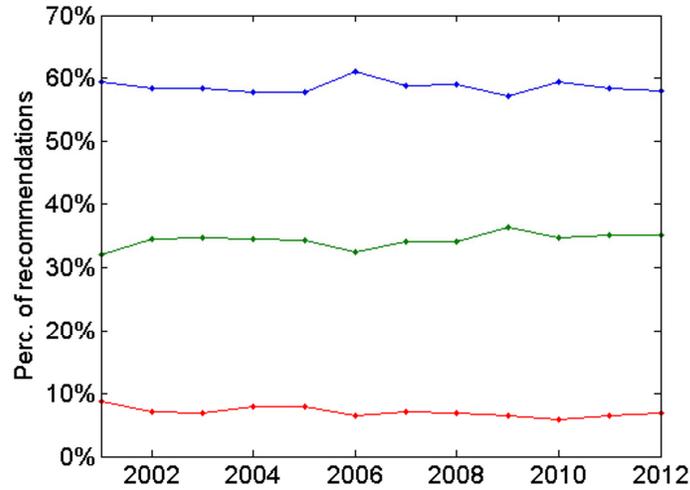

Figure 2. Percentage of recommendations with a score of 1 ('good'; shown in blue), 2 ('very good'; shown in green), or 3 ('exceptional'; shown in red) per year.

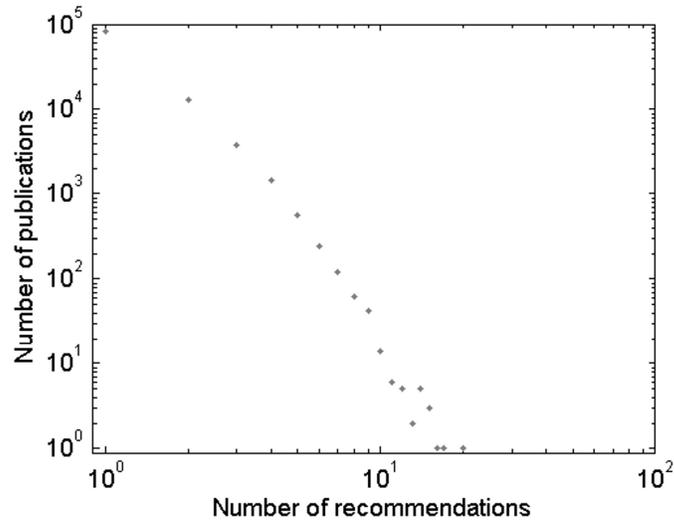

Figure 3. Distribution of the number of recommendations per publication. Notice that the horizontal and the vertical axis both have a logarithmic scale.

As already mentioned, the average number of recommendations per publication equals 1.30 (taking into account only publications that have been recommended at least once). Figure 3 shows the distribution of the number of recommendations per publication. The publication that has been recommended most has 20 recommendations.[2] Of the 102,360 publications that have been recommended, 81.1% have only one recommendation.

**3.2. Timing of recommendations**

In this subsection, we explore the timing of F1000 recommendations. For each recommendation, we know the month in which the recommendation was given as well as the month in which the recommended publication appeared. Our analysis focuses on the number of months between the appearance and the recommendation of a publication.

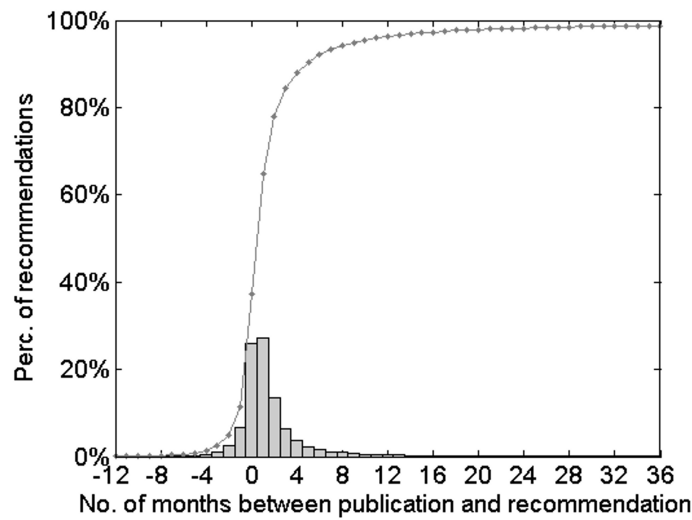

Figure 4. Distribution of recommendations by the number of months between publication and recommendation. The cumulative distribution is shown as well.

Figure 4 presents the distribution of recommendations by the number of months between publication and recommendation. We note that about 12% of the recommendations were given in an earlier month than the month in which the corresponding publication appeared. Apparently, F1000 faculty members sometimes recommend publications before their official date of appearance. Such early recommendations are probably caused by journals that make publications available online before actually assigning them to a journal issue or journals that release issues before their official publication date. What might also play a role is the availability of preprints in online repositories and perhaps the more informal exchange of manuscripts between authors and F1000 faculty members. The main difficulty is that we do not know when exactly a publication became publicly available. For this reason, the results presented in this subsection should be interpreted with some care.

---

[2] This is the following publication: Lolle, S.J., Victor, J.L., Young, J.M., & Pruitt, R.E. (2005). Genome-wide non-mendelian inheritance of extra-genomic information in Arabidopsis. *Nature*, *434*, 505–509.

The most important result from Figure 4 is that more than 80% of all recommendations are given between the second month before and the fourth month after the appearance of a publication. The month in which a publication appeared and the month thereafter together account for over 50% of all recommendations. Fewer than 10% of all recommendations are given six or more months after the appearance of a publication, and in fewer than 2% of all cases an F1000 faculty member is recommending a publication that is more than two years old. On the other hand, however, we note that there are a few recommendations that go back more than 50 years in time, referring to publications from the 1940s and 1950s.

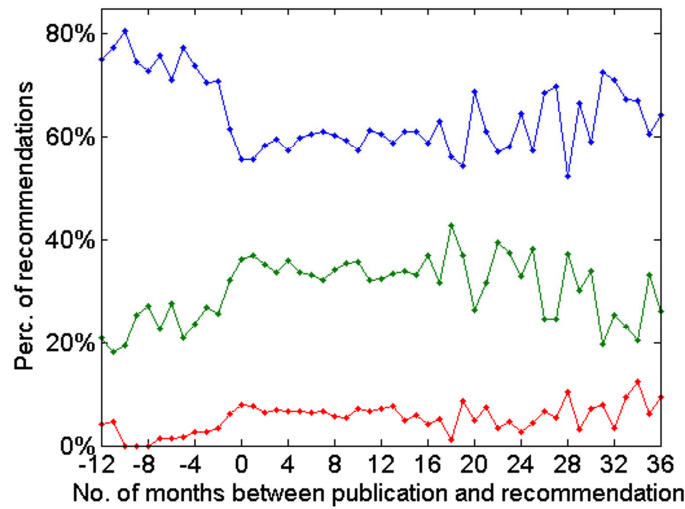

Figure 5. Percentage of recommendations with a score of 1 ('good'; shown in blue), 2 ('very good'; shown in green), or 3 ('exceptional'; shown in red) as a function of the number of months between publication and recommendation.

Figure 5 shows the proportion of recommendations with a score of 1, 2, or 3 as a function of the number of months between publication and recommendation. As can be seen in the figure, there does not exist a strong relation between the type of a recommendation and the timing of the recommendation. Recommendations of the 'good' type are overrepresented among the recommendations given two or more months before the official appearance of a publication, but it should be kept in mind that the number of such early recommendations is relatively small.

We now look at the way in which the number of months between publication and recommendation has changed over time. For each publication, we calculated both the average time to a recommendation and the time to the first recommendation. To ensure that earlier publication years can be compared with more recent ones in a valid way, we did not take into account recommendations given more than one year before or more than one year after the official appearance of a publication. Figure 6 shows both the average time to a recommendation and the average time to a publication's first recommendation as a function of the publication year. The number of months between publication and recommendation turns out to have been fairly stable over time, although there seems to be a small increasing trend.

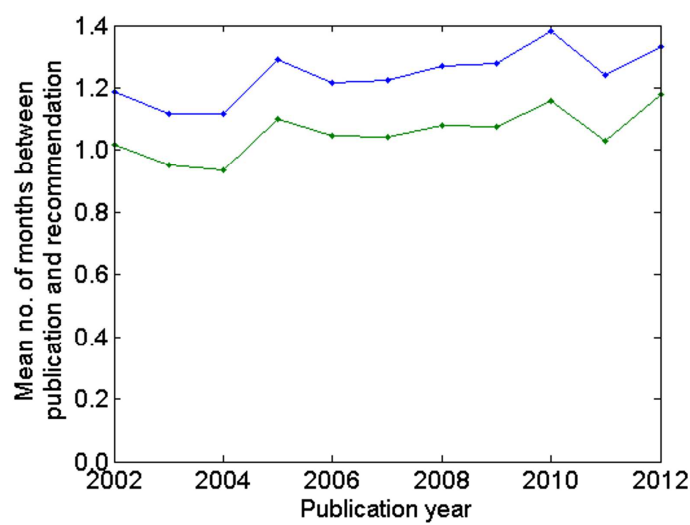

Figure 6. Average time to a recommendation (shown in blue) and average time to a publication's first recommendation (shown in green) as a function of the publication year.

**3.3. Recommendation activity per field of science**

The focus of F1000 is on research in biology and medicine. In this subsection, we examine how the recommendation activity of F1000 faculty members is distributed over different biological and medical fields. Our analysis relates to 38,327 publications from the period 2006–2009 that have at least one F1000 recommendation and that have been successfully matched with the WoS database. These are the same publications that will be considered in the comparison between recommendations and citations presented in the next subsection.

We use the journal subject categories in the WoS database to define fields of science.[3] For each subject category, the number of publications of the document types *article* and *review* in the period 2006–2009 was determined and the proportion of these publications with one or more recommendations was calculated. A fractional counting approach was taken in the case of publications belonging to multiple subject categories.

In the period 2006–2009, 172 of the 250 subject categories in the WoS database have at least one publication with a recommendation. It turns out that in some cases recommendations have been given to publications in subject categories that do not seem directly related to biology and medicine. Some examples of these subject categories are *Engineering, electrical & electronic*, *Information science & library science*, and *Sociology*. Each of these subject categories includes one or more recommended publications.

The 60 subject categories with the highest proportion of recommended publications in the period 2006–2009 are listed in Table A1 in the appendix. These 60 subject categories include almost 97% of all recommended publications in the period 2006–2009. As can be seen in Table A1, the *Multidisciplinary sciences* subject category has the highest proportion of recommended publications (11.8%). Given the presence of *Nature*, *PNAS*, and *Science* in this subject category, this is probably not a surprising finding. In addition to *Multidisciplinary sciences*, there are four other subject categories in which more than 5% of all publications have been recommended. These are *Developmental biology* (6.9%), *Anesthesiology* (6.4%), *Cell biology* (6.2%), and *Critical care medicine* (5.1%). The proportion of recommended publications in these subject categories is more than ten times as high as for instance in the *Surgery* subject category (0.5%). This seems to indicate that some biological and medical fields receive substantially more attention from F1000 faculty members than others.

---

[3] F1000 uses its own field classification system. This system is different from the WoS subject categories. The reason why we do not use the field classification system of F1000 is that this system only includes publications that have been recommended. Non-recommended publications are not included in the system and therefore do not have a field assignment.

**3.4. Comparison between recommendations and citations**

To what extent do F1000 recommendations correlate with citations? To answer this question, we study 1.7 million publications from the period 2006–2009, as explained in Section 2. Of these publications, 38,327 (2.2%) have at least one recommendation. The other publications have not been recommended by F1000 faculty members. On average, each publication has been cited 7.7 times.

We examine two ways in which recommendations and citations may relate to each other. On the one hand, we analyze the relation between the highest recommendation a publication has received and the number of citations of the publication. On the other hand, we analyze the relation between the total number of recommendations of a publication and its number of citations. In the latter case, no distinction is made between 'good', 'very good', and 'exceptional' recommendations.

In addition to comparisons between recommendations and citations, we also compare recommendations with journal citation scores (JCSs). As explained in Section 2, the JCS of a publication in journal X equals the average number of citations received by all publications in journal X in the period 2006–2009. The average JCS of the 1.7 million publications in our analysis equals 7.3.

*Citation distributions*

Based on their maximum recommendation score, publications can be classified into four sets: Publications that have not been recommended, publications with a maximum recommendation score of 1 ('good'), publications with a maximum recommendation score of 2 ('very good'), and publications with a maximum recommendation score of 3 ('exceptional'). For each of the four sets of publications, Figure 7 shows the cumulative citation distribution. The figure for instance indicates that about 80% of the publications without a recommendation have fewer than ten citations, while this is the case for less than 5% of the publications with a maximum recommendation score of 3. A clear pattern can be observed in Figure 7. Publications with a maximum recommendation score of 1 tend to be cited more frequently than publications that have not been recommended, publications with a maximum recommendation score of 2 tend to be cited more frequently than publications with a maximum recommendation score of 1, and publications with a maximum recommendation score of 3 tend to be cited more frequently than publications with a maximum recommendation score of 2. This indicates a clear correlation between a publication's maximum recommendation score and its number of citations.

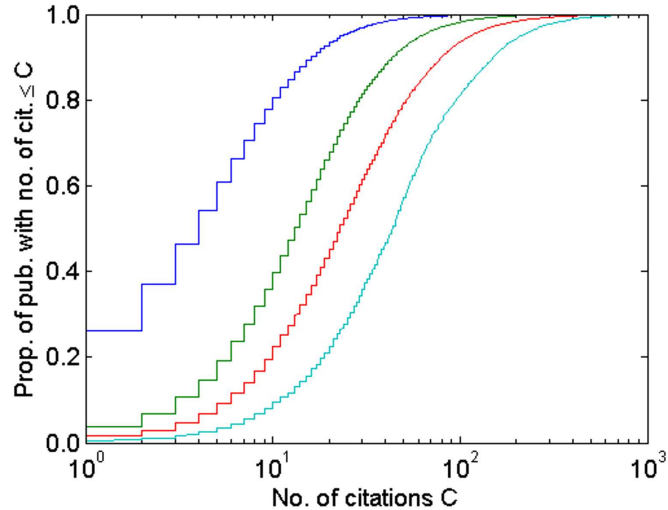

Figure 7. Cumulative citation distribution of publications with a maximum recommendation score of 0 (no recommendation; shown in blue), 1 ('good'; shown in green), 2 ('very good'; shown in red), or 3 ('exceptional'; shown in cyan). Notice that the horizontal axis has a logarithmic scale.

Figure 8 shows cumulative citation distributions for seven sets of publications defined based on the number of times publications have been recommended. The leftmost curve relates to publications that have not been recommended, the second curve from the left relates to publications that have been recommended once, the third curve from the left relates to publications that have been recommended twice, and so on. The rightmost curve relates to publications with six recommendations. Because there are only 101 publications with more than six recommendations, no citation distributions are shown for these publications. Like in Figure 7, a clear pattern is visible in Figure 8. Publications with more recommendations tend to receive more citations, although for publications with four, five, and six recommendations the difference seems to be relatively small. This suggests that, as the number of recommendations increases, the value of an additional recommendation becomes smaller.

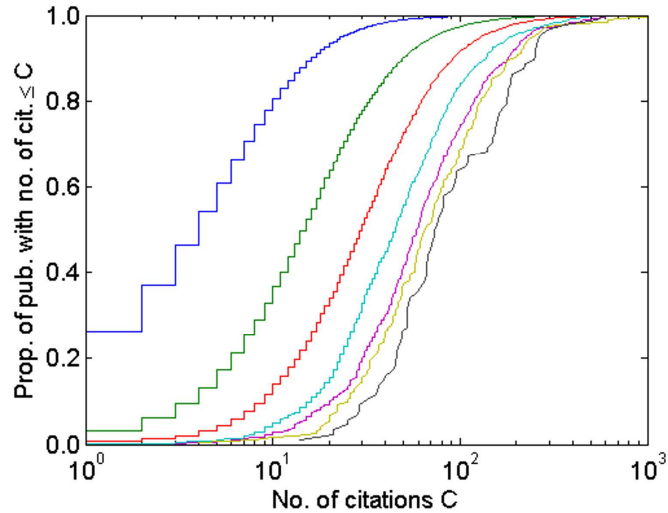

Figure 8. Cumulative citation distribution of publications with zero (leftmost curve) to six (rightmost curve) recommendations. Notice that the horizontal axis has a logarithmic scale.

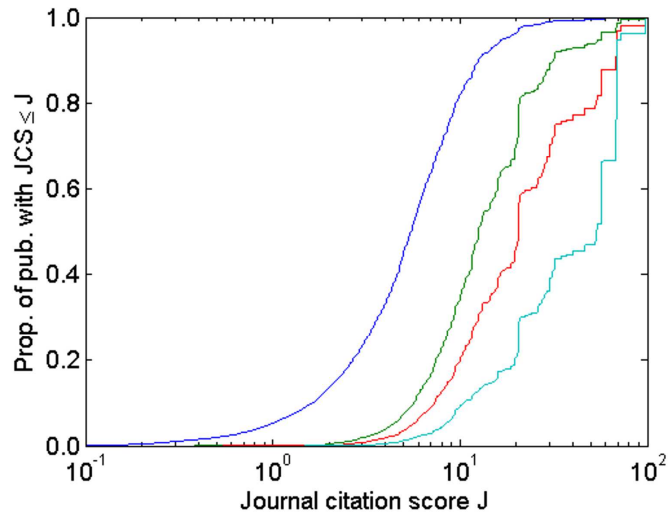

Figure 9. Cumulative JCS distribution of publications with a maximum recommendation score of 0 (no recommendation; shown in blue), 1 ('good'; shown in green), 2 ('very good'; shown in red), or 3 ('exceptional'; shown in cyan). Notice that the horizontal axis has a logarithmic scale.

Figures 9 and 10 are similar to Figures 7 and 8, but instead of citation distributions these figures show JCS distributions. The patterns visible in Figures 9 and 10 are similar to what we have observed in Figures 7 and 8. As can be seen in Figure 9, many publications with three or more recommendations have appeared in high-impact journals with a JCS above 50. The main journals in which these publications have appeared are *Nature*, *Science*, *Cell*, and *New England Journal of Medicine*.

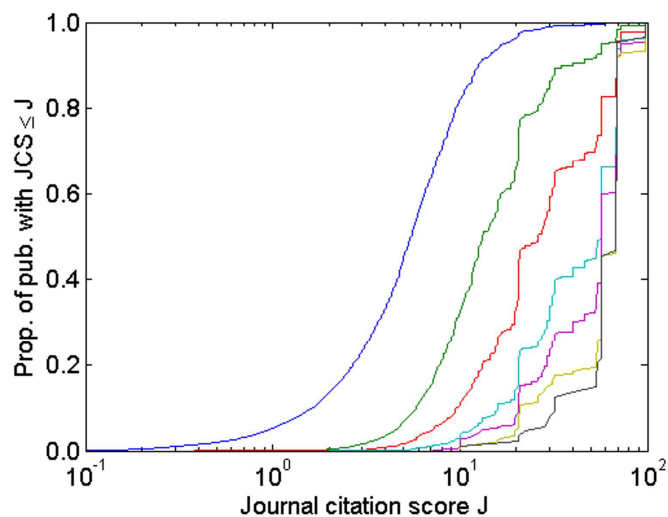

Figure 10. Cumulative JCS distribution of publications with zero (leftmost curve) to six (rightmost curve) recommendations. Notice that the horizontal axis has a logarithmic scale.

*Average citation scores*

Table 1 reports the average number of citations of publications with a maximum recommendation score of 0, 1, 2, or 3. The average JCS of these publications is reported as well. In addition, Table 1 also provides 95% confidence intervals. Like all confidence intervals reported in this paper, these confidence intervals were calculated using bootstrapping (e.g., Efron & Tibshirani, 1993).

Table 1. Average number of citations and average JCS of publications with a maximum recommendation score of 0 (no recommendation), 1 ('good'), 2 ('very good'), or 3 ('exceptional'). 95% confidence intervals are reported between brackets.

| Max. recommendation score | No. of publications | Mean no. of citations | Mean journal citation score |
|---|---|---|---|
| 0 | 1,669,304 | 7.2 [7.1, 7.2] | 6.9 [6.9, 7.0] |
| 1 | 22,862 | 20.7 [20.4, 21.1] | 17.4 [17.2, 17.6] |
| 2 | 12,838 | 37.6 [36.8, 38.6] | 27.9 [27.5, 28.3] |
| 3 | 2,627 | 68.6 [65.5, 72.3] | 44.6 [43.7, 45.6] |

In line with Figures 7 and 9, Table 1 indicates that both the average number of citations per publication and the average JCS per publication increase with the maximum recommendation score of a publication. The effect is quite strong, especially for the average number of citations per publication. Recall that on average the publications in our analysis have been cited 7.7 times. As can be seen in Table 1, publications that have not been recommended are somewhat below this average, publications with a maximum recommendation score of 1 are more than 2.5 times above the average, and publications with a maximum recommendation score of 2 are almost 5 times above the average. Publications with a maximum recommendation score of 3 even tend to be cited almost 9 times more frequently than the average.

Figures 11 and 12 show the relation between the number of recommendations of a publication and, respectively, the average number of citations and the average JCS. The figures also display 95% confidence intervals. Notice that for larger numbers of recommendations the confidence intervals are quite wide, especially in Figure 11. This is because there are only a relatively small number of publications that have been recommended more than a few times.[4]

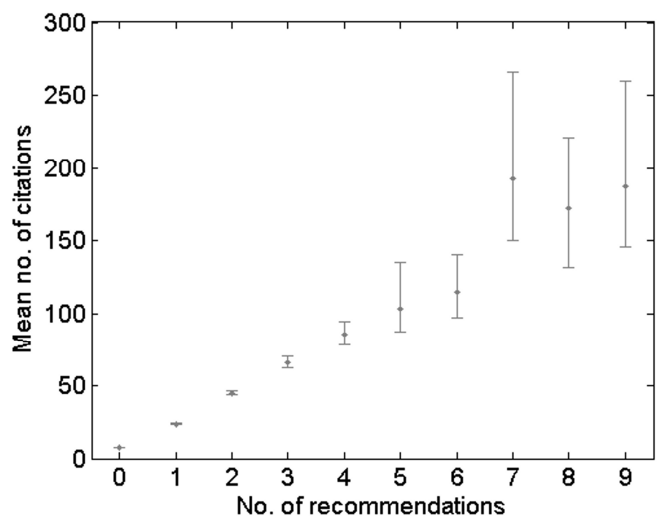

Figure 11. Relation between the number of recommendations of a publication and the average number of citations. The error bars indicate 95% confidence intervals.

---

[4] There are 12 publications with more than nine recommendations. These publications are not included in Figures 11 and 12.

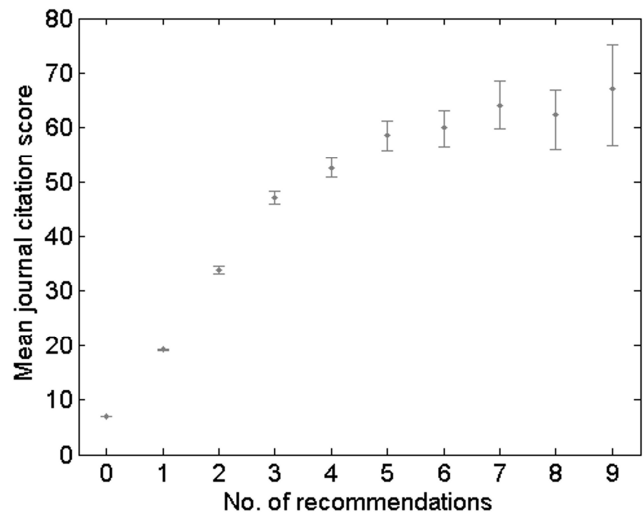

Figure 12. Relation between the number of recommendations of a publication and the average JCS. The error bars indicate 95% confidence intervals.

Both in Figure 11 and in Figure 12, a clear increasing trend can be observed. So in agreement with Figures 8 and 10, we find that on average publications with more recommendations also receive more citations and appear in journals with a higher citation impact. Notice in Figure 12 that for publications with three or more recommendations the effect of an additional recommendation on the average JCS is relatively small.

*Correlation coefficients*

Table 2 reports Pearson correlations between on the one hand publications' maximum recommendation score and number of recommendations and on the other hand publications' number of citations and JCS. Correlations obtained for the number of recommendations are slightly higher than those obtained for the maximum recommendation score, but the difference is very small. Since 97.8% of the publications in our analysis have not been recommended, it is not really surprising that the maximum recommendation score and the number of recommendations yield similar correlations. Notice that correlations of recommendations with the JCS are higher than correlations of recommendations with the number of citations. Hence, in terms of the Pearson correlation, recommendations are more strongly related to the citation impact of the journal in which a publication has appeared than to the number of citations of a publication.[5]

Table 2. Pearson correlations between on the one hand publications' maximum recommendation score and number of recommendations (either weighted or unweighted) and on the other hand publications' number of citations and JCS. 95% confidence intervals are reported between brackets.

|  | No. of citations | Journal citation score |
|---|---|---|
| Max. recommendation score | 0.24 [0.23, 0.26] | 0.33 [0.33, 0.34] |
| No. of recommendations | 0.26 [0.24, 0.28] | 0.34 [0.33, 0.34] |
| Weighted no. of recommendations | 0.27 [0.25, 0.29] | 0.34 [0.34, 0.35] |

So far, when counting the number of recommendations of a publication, we have given equal weight to 'good', 'very good', and 'exceptional' recommendations. A better approach may be to give different weights to these different types of recommendations. One way in which the weights could be determined is by choosing them in such a way that the Pearson correlation between on the one hand the weighted number of recommendations of publications and on the other hand either the number of citations or the JCS of publications is maximized. This amounts to performing a least-squares linear regression with the number of citations or the JCS as the dependent variable and the number of 'good', 'very good', and 'exceptional' recommendations as independent variables.

Table 3 reports the results of the linear regressions, with $\beta_1$, $\beta_2$, and $\beta_3$ denoting the regression coefficients for, respectively, the number of 'good', 'very good', and 'exceptional' recommendations and $\alpha$ denoting the intercept. Using the number of citations as the dependent variable, we find that a 'very good' recommendation should be given about $\beta_2 / \beta_1 = 1.8$ times as much weight as a 'good' recommendation, while an 'exceptional' recommendation should be given about $\beta_3 / \beta_1 = 2.5$ times as much weight.[6] Considerably smaller weight differences are obtained when instead of the

---

[5] We also tested the effect of applying a logarithmic transformation to the number of citations of a publication. This turned out to yield lower correlations between recommendations and citations than the ones reported in Table 2.

[6] These weights are fairly close to the weights used by F1000 to calculate the total recommendation score of a publication. F1000 assigns weights of 1, 2, and 3 to, respectively, 'good', 'very good', and

number of citations the JCS is used as the dependent variable. The correlations obtained by weighting recommendations based on the regression coefficients reported in Table 3 are shown in the last row of Table 2. As can be seen, these correlations are only marginally higher than the correlations obtained without weighting. This is a consequence of the dominance of our analysis by publications without recommendations. For these publications, giving different weights to different types of recommendations does not make any difference.

Table 3. Results of least-squares linear regressions with the number of citations or the JCS as the dependent variable and the number of 'good', 'very good', and 'exceptional' recommendations as independent variables. 95% confidence intervals are reported between brackets.

| Regression coefficient | Dependent variable | |
|---|---|---|
| | No. of citations | Journal citation score |
| $\beta_1$ | 12.2 [11.8, 12.7] | 9.3 [9.2, 9.6] |
| $\beta_2$ | 22.4 [21.5, 23.4] | 14.1 [13.8, 14.4] |
| $\beta_3$ | 31.1 [27.9, 34.4] | 15.0 [13.8, 16.1] |
| $\alpha$ | 7.2 [7.1, 7.2] | 7.0 [7.0, 7.0] |

As we have seen in Subsection 3.2, recommendations are mainly given in the first few months after a publication has appeared. Because of this, one may consider to use the recommendations received by a publication as a predictor of the number of citations the publication will receive. From this point of view, recommendations can be seen as an alternative to the citation impact of the journal in which a publication has appeared, since journal citation impact is also often interpreted as a predictor of the number of citations of a publication.

An obvious question is whether for the purpose of predicting the number of citations of a publication recommendations may be more accurate than journal citation impact. Based on the Pearson correlation, the answer to this question is negative. The Pearson correlation between publications' JCS and their number of

---

'exceptional' recommendations, and it calculates the total recommendation score of a publication as the sum of the weights of the recommendations given to the publication. In our analysis, we work with the weights obtained from Table 3, but we also tested the effect of working with the weights used by F1000 and this turned out to yield virtually identical results.

citations equals 0.52.[7] This is much higher than the correlations between recommendations and citations reported in Table 2. Hence, according to the Pearson correlation, predictions of citations based on JCSs are substantially more accurate than predictions of citations based on recommendations.[8]

*Highly cited publications*

Given the fact that 97.8% of the publications in our analysis have not been recommended at all, it is perhaps not surprising that the correlation between recommendations and citations is much weaker than the correlation between JCSs and citations. Because of the low percentage of publications with recommendations, one could hypothesize that recommendations mainly indicate the most highly cited publications in the scientific literature. To test this idea, we identified the top 1% most highly cited publications (i.e., all publications with at least 58 citations) among the 1.7 million publications included in our analysis. We then examined to what extent recommendations and JCSs are able to distinguish between these highly cited publications and the other 99% of the publications.

Figure 13 presents precision-recall curves obtained for four approaches for identifying the top 1% most highly cited publications in our analysis. For a given selection of publications, precision is defined as the number of highly cited publications in the selection divided by the total number of publications in the selection. Recall is defined as the number of highly cited publications in the selection divided by the total number of highly cited publications. Of the four approaches for identifying highly cited publications that are considered in Figure 13, one is based on JCSs (shown in blue) and the other three are based on recommendations. The latter approaches identify highly cited publications based on a publication's maximum recommendation score (shown in green), a publication's unweighted number of recommendations (shown in red), or a publication's weighted number of recommendations, with weights obtained from Table 3 (shown in cyan). All four approaches deal with ties (e.g., multiple publications with the same JCS or the same maximum recommendation score) by selecting publications in random order.

---

[7] Strictly speaking, the correlation coefficient of 0.52 is not a valid measure of the degree to which the JCS of a publication *predicts* the number of citations of the publication. To measure the predictive power of JCSs, we should calculate JCSs based on publications from an earlier time period. Using JCSs calculated based on publications from the period 2002–2005, a correlation coefficient of 0.49 instead of 0.52 is obtained. Given the small difference, we simply use JCSs calculated based on publications from the period 2006–2009 in our analysis. This has the advantage that we avoid the complexity of having two different sets of JCSs.

[8] Using least-squares linear regression, a combined predictor based on both the JCS of a publication and the number of 'good', 'very good', and 'exceptional' recommendations of a publication can be constructed. This combined predictor has a Pearson correlation of 0.53 with citations, which indicates that combining journal citation impact with recommendations has hardly any added value compared with the use of journal citation impact only.

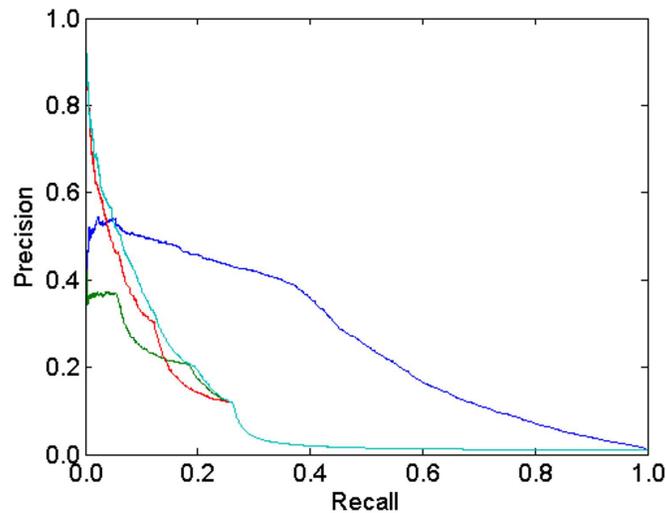

Figure 13. Precision-recall curves for four approaches for identifying the top 1% most highly cited publications. The approaches are based on the JCS of a publication (blue curve), the maximum recommendation score of a publication (green curve), the unweighted number of recommendations of a publication (red curve), and the weighted number of recommendations of a publication (cyan curve).

To illustrate the interpretation of the precision-recall curves in Figure 13, we take the curve obtained based on publications' maximum recommendation score as an example. This curve for instance indicates that a recall of 0.10 (or 10%) corresponds with a precision of 0.25 (or 25%). What does this mean? To see this, suppose we select a certain number of publications, where the selection is made based on publications' maximum recommendation score. A recall of 0.10 combined with a precision of 0.25 then means that, if we want 25% of the publications in our selection to belong to the top 1% most highly cited, our selection can include only 10% of all top 1% most highly cited publications. The curve also indicates that a recall of 0.20 (or 20%) corresponds with a precision of 0.17 (or 17%). This means that, if we are satisfied with only 17% of the publications in our selection belonging to the top 1% most highly cited, it becomes possible to include 20% of all top 1% most highly cited publications in our selection.

The main conclusion that we can draw from Figure 13 is that JCSs perform much better than recommendations for the purpose of identifying the top 1% most highly cited publications in our analysis. Only at very low levels of recall, the weighted and unweighted number of recommendations yield a higher precision than the JCS. The maximum recommendation score is always outperformed by the JCS. The relatively low precision/recall values obtained using recommendations can be explained by the fact that 73.7% of the top 1% most highly cited publications have not been recommended at all. Results similar to those presented in Figure 13 are obtained when instead of the top 1% most highly cited publications we consider the top 0.1% or the top 10%. One of the results we obtain is that about half of the recommended publications belong to the top 10% most highly cited publications. The other half of the recommended publications belong to the bottom 90% in terms of citations. Based on the results of our precision-recall analysis, we conclude that JCSs are substantially more accurate than recommendations not only for predicting citations in general but also for the more specific task of predicting the most highly cited publications.

*Sensitivity analyses*

It should be noted that all results presented in this subsection could be sensitive to the selection of publications included in our analysis. We therefore also calculated results based on a different selection of publications. Instead of selecting 1.7 million publications from 5,908 'microfields' with at least one F1000 recommended publication (see Section 2), we selected 1.1 million publications from 3,044 microfields with at least three F1000 recommended publications. The results turned out to be similar to the ones presented above, indicating that the results of our analysis have only a limited sensitivity to the selection of publications.

Another type of sensitivity analysis was suggested to us by F1000. It may be that high-impact journals are different from ordinary journals in terms of both citation and recommendation characteristics. Recommendations may therefore be especially suitable for identifying highly cited publications in low- and medium-impact journals. We tested this hypothesis by excluding from our analysis all publications in journals with a JCS above 50 (e.g., *Nature*, *Science*, *Cell*, and *New England Journal of*

*Medicine*). This lead to the exclusion of about 0.9% of the 1.7 million publications included in our original analysis. Of the 38,327 publications with one or more recommendations, 14.4% were excluded. Using the non-excluded publications, we performed a precision-recall analysis similar to the one reported above. This yielded results that are substantially worse than the ones presented in Figure 13. Precision/recall values obtained using both JCSs and recommendations are considerably below the values obtained in the original analysis, although JCSs still outperform recommendations. Based on this outcome, it can be concluded that leaving out publications in high-impact journals does not improve the ability of recommendations to identify highly cited publications.

**3.5. Comparison between recommendations and field-normalized citations**

It is well-known that citation behavior differs widely across fields of science. As a consequence, publications in one field may on average receive substantially more citations than publications in another field. In medical research, for instance, there seems to be a tendency for publications in basic fields to be cited more frequently than publications in clinical fields (Seglen, 1997; Van Eck, Waltman, Van Raan, Klautz, & Peul, 2012). In the previous subsection, we did not take into account the issue of differences in citation behavior between fields of science. We simply assumed citations in one field to be directly comparable to citations in another field. We now examine to what extent taking into account the issue of between-field differences in citation behavior may lead to different results.

For each publication, we calculated a field-normalized citation score by dividing the number of citations of the publication by the average number of citations of all publications in the same 'microfield' in the period 2006–2009 (see Section 2). Next, for each journal, we calculated a field-normalized JCS by averaging the normalized citation scores of all publications of the journal. In the terminology of Waltman, Van Eck, Van Leeuwen, Visser, & Van Raan (2011), we calculated each journal's mean normalized citation score. We then identified for each microfield separately the top 1% most highly cited publications. Like at the end of the previous subsection, we are interested in the degree to which F1000 recommendations and JCSs are able to identify highly cited publications. However, unlike in the previous subsection, highly cited publications are defined locally per microfield rather than globally for all microfields together. Also, instead of ordinary JCSs, we use field-normalized JCSs.

Like Figure 13 in the previous subsection, Figure 14 presents precision-recall curves obtained for four approaches for identifying highly cited publications. The difference with Figure 13 is that highly cited publications are defined locally per microfield and that JCSs are field normalized. Our expectation was that correcting for differences in citation behavior between fields would lead to improved results in terms of precision and recall, at least when recommendations are used to identify highly cited publications. However, comparing Figure 14 with Figure 13, it can be seen that the results have worsened rather than improved. At any level of recall, all four approaches for identifying highly cited publications have a lower level of precision. We also find that only 20.1% of the top 1% most highly cited publications have been recommended at least once, while in the previous subsection this was the case for 26.3% of the top 1% most highly cited publications.

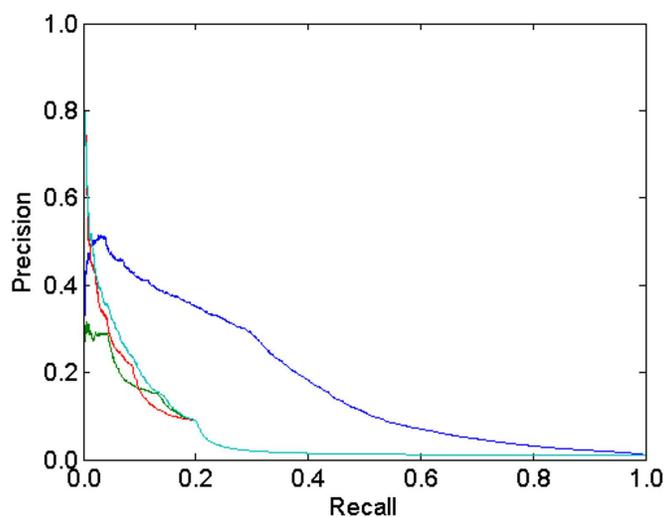

Figure 14. Precision-recall curves for four approaches for identifying the top 1% most highly cited publications per microfield. The approaches are based on the field-normalized JCS of a publication (blue curve), the maximum recommendation score of a publication (green curve), the unweighted number of recommendations of a publication (red curve), and the weighted number of recommendations of a publication (cyan curve).

Why does correcting for between-field differences in citation behavior lead to worse results? Our idea is that this is probably due to the uneven distribution of recommendation activity over fields, as discussed in Subsection 3.3. Looking at Table

A1 in the appendix, we observe that two fields with a lot of recommendations, both in absolute and in relative terms, are *Cell biology* and *Biochemistry & molecular biology*. These fields are well known as fields with a high citation density (i.e., a large average number of citations per publication). This suggests that to some extent having a high recommendation activity may be correlated with having a high citation density. Such a correlation between recommendation activity and citation density would explain why field normalization of citations weakens the relation between recommendations and citations.

## 4. Discussion and conclusion

Our large-scale analysis of F1000 recommendations indicates that about 2% of the publications in the biological and medical sciences receive one or more recommendations from F1000 faculty members. The exact percentage depends on how one chooses to delineate the biomedical literature. If a publication is recommended, the number of recommendations is usually small, with an average of 1.30 recommendations per publication. Most recommendations are given shortly after (or sometimes shortly before) the official date at which a publication appeared. Over 90% of all recommendations are given before the sixth month after a publication's appearance. Our link between F1000 recommendations and publications in the Web of Science bibliographic database suggests that the proportion of recommended publications differs quite substantially across fields, with publications in the field of cell biology for instance being more than ten times as likely to be recommended as publications in the field of surgery.

In line with earlier studies based on smaller data sets (Bornmann & Leydesdorff, 2013; Li & Thelwall, 2012; Medical Research Council, 2009; Priem et al., 2012; Wardle, 2010), our analysis shows a clear correlation between F1000 recommendations and citations. How should we qualify the strength of this correlation? The answer to this question very much depends on the point of view one takes. In our view, the best way to answer this question is to compare the correlation between recommendations and citations with the correlation between journal citation scores and citations. Journal citation scores, for instance journal impact factors, are often regarded as fairly weak predictors of citation scores at the level of individual publications (e.g., Seglen, 1997). Our analysis indicates that the correlation between recommendations and citations is considerably weaker than the correlation between journal citation scores and citations. So if journal citation scores are judged to be fairly weak predictors of publication citation scores, this judgment should definitely extend to recommendations as well.

In a sense, F1000 recommendations cannot be expected to correlate very strongly with citations, simply because about 98% of all biomedical publications do not have any recommendation at all. A more reasonable idea may be that recommendations predict highly cited publications. Our analysis shows that also from this point of view recommendations have a lower predictive power than journal citation scores. It turns out that journal citation scores are substantially more accurate predictors of high citedness than recommendations. On the one hand, we do find that recommended publications tend to be cited quite a lot, with for instance half of the recommended publications belonging to the top 10% most highly cited publications in our analysis. On the other hand, however, we also find that many highly cited publications have not been recommended. For instance, almost three-quarter of the top 1% most highly cited publications have not been recommended. Our analysis also indicates that correcting for differences in citation behavior between fields does not increase the predictive power of recommendations.

From the research evaluation perspective, how should one interpret the relatively weak correlation between F1000 recommendations and citations? On the one hand, one could interpret this as an indication that, contrary to what it claims, F1000 fails to consistently identify the most important publications in the biological and medical sciences. This would be in line with the conclusion drawn by Wardle (2010) for the field of ecology. Wardle argues that in the field of ecology F1000 recommendations are a poor predictor of highly cited publications and suggests this to be caused by the uneven distribution of F1000 faculty members over different areas of ecological research, the problem of cronyism, and the problem of geographical bias.

However, the relatively weak correlation between recommendations and citations could also be interpreted in a different way. It could be argued that recommendations and citations simply do not capture the same type of impact. This is similar to the reasoning of Li and Thelwall (2012), who suggest that recommendations measure the 'quality of articles from an expert point of view' while citations measure 'research impact from an author point of view'. Following this reasoning, one would expect F1000 recommendations to sometimes identify important publications that remain unnoticed by citation analysis. The other way around, this reasoning might explain why some highly cited publications are not recommended.

Based on our analysis, which of the above two interpretations is more valid cannot be established. This would require a more in-depth investigation of, for instance, the reasons F1000 faculty members have to recommend a publication, but perhaps also of possible biases in F1000's peer-nomination system for selecting faculty members (as suggested by Wardle, 2010). These topics may be worthwhile to investigate in future studies.

## Acknowledgment

We would like to thank Omer Gazit, Jane Hunter, Ian Tarr, and Kathleen Wets from F1000 for providing us with the F1000 data set, for helping us with the interpretation of the data, and for providing feedback on an earlier version of this paper. The views expressed in this paper are those of the authors and are not necessarily shared by F1000.

## Appendix

Table A1. Top 60 Web of Science journal subject categories with the highest percentage of publications with one or more recommendations. For each subject category, both the number and the percentage of recommended publications are reported.

| | | | | | |
|---|---|---|---|---|---|
| Multidisciplinary sciences | 5895 | 11.8% | Allergy | 73 | 1.5% |
| Developmental biology | 702 | 6.9% | Obstetrics & gynecology | 395 | 1.5% |
| Anesthesiology | 732 | 6.4% | Psychiatry | 428 | 1.4% |
| Cell biology | 3051 | 6.2% | Parasitology | 133 | 1.3% |
| Critical care medicine | 468 | 5.1% | Oncology | 983 | 1.3% |
| Immunology | 1575 | 3.5% | Biophysics | 276 | 1.2% |
| Genetics & heredity | 1205 | 3.1% | Math. & comp. biology | 79 | 1.2% |
| Biochem. & molecular biology | 3601 | 3.1% | Plant sciences | 482 | 1.0% |
| Cell & tissue engineering | 43 | 3.1% | Neuroimaging | 25 | 1.0% |
| Neurosciences | 2373 | 3.0% | Andrology | 12 | 0.9% |
| Urology & nephrology | 1005 | 2.9% | Biochemical research methods | 227 | 0.9% |
| Dermatology | 490 | 2.6% | Pediatrics | 289 | 0.9% |
| Hematology | 642 | 2.5% | Pathology | 148 | 0.8% |
| Medicine, research & exp. | 746 | 2.5% | Chemistry, medicinal | 168 | 0.8% |
| Virology | 400 | 2.4% | Biotech. & appl. microbiology | 323 | 0.8% |
| Respiratory system | 349 | 2.2% | Biodiversity conservation | 38 | 0.7% |
| Rheumatology | 324 | 2.2% | Transplantation | 54 | 0.7% |
| Evolutionary biology | 194 | 2.1% | Emergency medicine | 47 | 0.7% |
| Gastroenterology & hepatology | 737 | 2.1% | Behavioral sciences | 49 | 0.6% |
| Microbiology | 837 | 2.0% | Geriatrics & gerontology | 46 | 0.6% |
| Vascular diseases | 446 | 2.0% | Psychology, clinical | 75 | 0.6% |
| Endocrinology & metabolism | 809 | 1.9% | Public, env. & occup. health | 268 | 0.6% |
| Medicine, general & internal | 1135 | 1.9% | Chemistry, multidisciplinary | 572 | 0.6% |
| Biology | 325 | 1.7% | Pharmacology & pharmacy | 427 | 0.5% |
| Reproductive biology | 129 | 1.6% | Surgery | 371 | 0.5% |
| Infectious diseases | 323 | 1.6% | Tropical medicine | 27 | 0.5% |
| Ecology | 470 | 1.6% | Toxicology | 90 | 0.5% |
| Physiology | 329 | 1.6% | Psychology, multidisciplinary | 128 | 0.5% |
| Cardiac & cardiovas. systems | 688 | 1.5% | Social sciences, biomedical | 14 | 0.4% |
| Clinical neurology | 709 | 1.5% | Health care sciences & serv. | 50 | 0.4% |